\documentclass[twocolumn,pre,superscriptaddress,letterpaper]{revtex4} 
\usepackage{graphicx}
\usepackage{dcolumn}

\def\e{\epsilon}

\newcommand{\be}{\begin{equation}} 
\newcommand{\ee}{\end{equation}} 
\newcommand{\bd}{\begin{displaymath}} 
\newcommand{\ed}{\end{displaymath}} 
\newcommand{\bea}{\begin{eqnarray}} 
\newcommand{\eea}{\end{eqnarray}} 
\newcommand{\beay}{\begin{eqnarray*}} 
\newcommand{\eeay}{\end{eqnarray*}} 
\newcommand{\bc}{\begin{center}} 
\newcommand{\ec}{\end{center}}

\newcommand{\noi}{\noindent}
 
\begin{document} 
\title{A Thermodynamic Model for Prebiotic Protein Function}

\author{Ay\c se Erzan}
\affiliation{Department of Physics, Faculty of Sciences and 
Letters\\ Istanbul Technical University, Maslak 80626, Istanbul, Turkey}
\affiliation{G\"ursey Institute, P.O.B. 6, \c Cengelk\"oy, 81220 Istanbul, Turkey }
\author{Erkan T\"uzel\footnote[2]{Present address: University of Minnesota, School of Physics and Astronomy, 
116 Church St. SE, Minneapolis, MN, 55455, USA}}
\affiliation{Department of Physics, Faculty of Sciences and 
Letters\\ Istanbul Technical University, Maslak 80626, Istanbul, Turkey}

\date{\today}
\begin{abstract}                     
We propose a scenario for the prebiotic co-evolution of RNA and of
fast folding proteins with large entropy gaps as observed today.
We show from very general principles that the folding and unfolding of the proteins
synthesized by RNA can function as a heat pump. Rock surfaces can facilitate the folding of 
amino acid chains having polar and hydrophobic residues, with an accompanying heat loss to the 
surrounding rock. These chains then absorb heat from the soup as they unfold. This
opens the way to the enhancement of RNA replication rates, by the enzymatic action of folded
proteins present in greater numbers at reduced temperatures. This gives an evolutionary advantage
to those RNA coding amino acid sequences with non-degenerate folded states which would provide the
most efficient refrigeration.
 
\end{abstract}
\maketitle
\section{Introduction}

In order to understand the process of selection that could have 
lead to  the observed properties of biological proteins, it is 
natural to look for mechanisms whereby proteins, which cannot 
duplicate themselves, evolved together with 
self-replicating molecules, 
``pre-living''concentrations of RNA,
which provide the code for protein 
synthesis.~\cite{eigen,Eigen1,gesteland,selfish,Smith1,Smith99,Kauffmann} 
The enzymatic action of proteins on the self-replication of RNA 
constitutes a hypercyle,  which may be considered the elementary 
unit of  
evolution~\cite{eigen,Eigen1}.
  
Biological proteins fulfill their  functions 
in unique folded states, which they are able to reach in a very 
short time after being synthesized.~\cite{protfold}
Small single domain proteins fold into 
their secondary structures within milliseconds~\cite{fersht}, or 
even faster~\cite{mayor}.
 These ``native" states correspond to  minimum  
(free) energy configurations.~\cite{Wolynes} 
A random sequence of amino acids, however, will typically have a 
degenerate ground state.~\cite{Shakhnovich1,Shakhnovich2}
We would like to propose a possible pathway in which those amino-acid sequences with unique ground states were selected  in the course of evolution, without begging the question by referring to highly evolved biological functions.
We will assume that this selection must have occurred in the prebiotic soup, where local temperature differences could have a large effect on  the efficiency of RNA replication.

As a chain  folds (unfolds) at constant ambient temperature, heat 
will be given off (absorbed) by an amount proportional to the 
difference in entropy between the unfolded and folded states. 
Since we are interested in relatively high temperatures, and our
proteins do not fold into their tertiary structures, we will ignore
their hydrational entropy changes.
For an amino acid chain of a given length,  
this difference will be the largest if the folded state corresponds to 
a non-degenerate ground state, or several possible low-energy configurations 
that are well isolated from each other by very high free energy barriers.
We will argue that amino acid chains which essentially behave as  two-state systems~\cite{gesteland,kaya}
with large entropy gaps prove to be the most effective refrigerants, if they were to 
be employed in a refrigeration cycle. Selection of amino-acid chains with folded 
states in deep free-energy wells could then be succeeded by the the evolution of 
more specific functions, leading to the pruning of those low-lying states so 
that only one, serving a highly specialized enzymatic activity, would survive.
For convenience, we will henceforth use the term ``proteins" to mean amino acid 
chains, regardless of their degree of evolution. 

It has been pointed out~\cite{Eigen1} that a rudimentary form of compartition is necessary for evolutionary processes to be possible, and porous rock~\cite{Smith99,Kauffmann} is among the likely environments  to have 
played host to prebiotic processes, in a ``soup,''consisting  of both organic and inorganic 
materials, that are assumed to 
be present in the prebiotic earth. 
For simplicity, 
let us concentrate on a series of compartments that contain 
RNA, protein molecules, amino acids and water. The 
surrounding rock is bathed by water at some medium temperature.  
The temperature range we  have in mind is such that the RNA and 
proteins are stable, and will vary, say, between 300-360 K, (the 
denaturation temperatures of most proteins are in this 
range~\cite{fersht}).

In the next section, Section II, we will describe how proteins 
could act as a refrigerant in an adsorption refrigerator. 
In Section III, we discuss a toy Hamiltonian for 
the guided and unguided folding of chains, and describe the refrigeration cycle in the entropy-temperature plane.
A discussion and 
pointers for 
further research are provided in the last section.

\section{Proteins in an adsorption refrigeration cycle}

The hydrophobic interactions which drive the folding of proteins into their native states~\cite{hydrophob,Burak,Widom} 
may also  make it 
favorable for non-polar residues on the chain to adsorb on 
nearby hydrophobic surfaces~\cite{Pinar} lowering the free energy of the 
whole system by reducing the number of water 
molecules in interaction with the non-polar residues. If we assume that a 
certain fraction of the non-polar residues have adsorbed on the surface, 
in a relatively stretched conformation(as shown in Fig.1),  
then it can 
slightly  increase the entropy with the diffusion and 
aggregation of the non-polar residues on the surface, allowing the 
intervening sections of the chain greater freedom. The residues in the ``loops`` 
are then free to fold into ``beads'' or ``droplets'' that are the incipient 
building blocks of the secondary structure.

If this partially folded state is on 
the ``correct pathway''~\cite{Levinthal,bakk1,hansen} to a low energy folded 
conformation of the chain, stabilized by the specific intra-chain 
interactions for the  given sequence of amino-acids,
the rock surface can be said to act as a guide for the folding process.

\begin{figure}
\bc
\leavevmode
\includegraphics[width=9cm,angle=0]{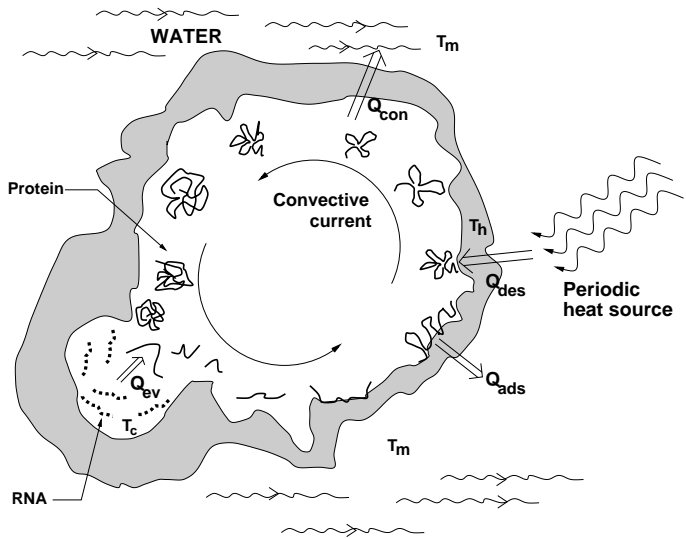}
\caption{\small{
Cartoon showing the crossection of a pore, with the different steps in an
adsorption--refrigeration cycle involving denatured and folded proteins.  
}}\label{kayapro}
\ec
\end{figure}

We would like to propose that the relatively large heat of adsorption and denaturation which can be achieved by proteins may make them amenable to functioning as refrigerants in an adsorption refrigerator.

Unlike work-driven refrigerators, the conventional adsorption refrigerator~\cite{Cengel,Hsieh,Teng,Meunier,Pons} relies on the availability of a cheap heat source and has no moving parts.  The refrigerant absorbes heat from the cold reservoir at $T_c$, as it flash evaporates, and then is led to an adsorber bed (rather than a compressor), where it adsorbes on a substrate at some medium temperature $T_m$, where it gives off a heat of adsorption $Q_{\rm ads}$ to the environment.  To regenerate the refrigerant, the ``low quality'' heat source is used to heat the adsorber bed to a high temperature, $T_h$. The refrigerant that is released in this way is led to a condenser (which may be again at $T_m$), liquified, led through a nozzle to lower the boiling temperature, and then piped once more to the evaporation chamber.  A ``batch operating'' system uses two ``beds''(as adsorber and desorber), the cycle switching between them in turn.~\cite{Chua}

The ``gas'' and ``liquid'' phases correspond, in our case,  to the unfolded and folded states of the protein chains. The physical environment we have in mind is a series of interconnected chambers in porous rock bathed on the outside with ambient water.  Part of the system is periodically exposed to intense heat. This may, for example be due to insolation or to the heat relased from a vent in the ocean floor.~\cite{Martin+Russel}

The refrigeration cycle consists of the following steps.

{\it i)}  The unfolded protein adsorbs on to a pore surface, where it folds.  

We assume the folding to take place isothermally,  at  the temperature  of  the rock 
surface, which we take to be $T_m$.  An amount of heat $Q_{\rm ads}$ will be given off in this process.

{\it ii)} The regenerative step involves the heating of the surface to dislodge the folded proteins from the wall.  This, we propose, may be supplied by sunshine or
 some geothermal source. During the regenerative step, a quantity of heat $Q_{\rm des}$ will be absorbed.

The chain can now detach from the rock, and 
complete its folding around its hydrophobic core, with the 
polar residues predominantly on the 
``outside.''~\cite{tuzel1,tuzel2} 
We have shown~\cite{Pinar}, using a lattice model for the hydrophobic interactions, that above the temperature interval where the hydrophobic interactions make it favorable for a hydrophobic chain to adsorb onto a hydrophobic boundary, there is a temperature interval where the chain prefers to detach from the wall and go into a folded state. Raising the temperature further results in the unfolding of the chains.

{\it iii)} These proteins (denatured or otherwise) shall now be convected
away from the hot wall.  More temperate regions of the porous network,
cooled by the ambient water, will act as the ``condenser'' in this system,
and as they cool to the ambient temperature $T_m$ once more, the proteins
will equilibrate to the native, folded states.  A quantity of heat $Q_{\rm
con}$ is given off in the process.

{\it iv)}  In order for the denaturation temperature to be lowered as the proteins enter the cool compartment at $T_c$, (analogously to passing them through a nozzle and lowering the pressure, in a conventional refrigerator), here use is made of the fact that the denaturation temperature depends very sensitively on the total ionization strength.  The gradient of ionization strength can, in fact, be easily achieved; the RNA molecules, the amino acids and denatured proteins in the interior of the chamber will lower the unfolding temperature for the coming chains.

{\it v)} As the proteins unfold in the cool chamber due to the lowered
denaturation temperature, they will absorb $Q_{\rm ev}$.  This phase of the
cycle corresponds to the ``evaporator'' of the adsorption refrigerator.

Clearly, for our adsorption refrigerator  to work, the adsorber ``beds'' must be swept clean not just of proteins, but other unreacted amino acids, or other hydrophobes.  The periodic nature of the heating serves this purpose as well.

This refrigeration cycle is similar to magnetic cooling, if one
thinks of the the action of the rock surface, facilitating the folding,  as the magnetic field. Note that for the chain to fold, and to
give off heat to the medium temperature reservoir, work has been done on it
by the combined action of the rock surface (entropy mediated hydrophobic
interactions) and intra-chain interactions. The ambient water bathing the outer surface of the rock acts as  the thermal contact.  

\section{A Hierarchical Two-state Model For Protein Folding}

To investigate the refrigeration 
cycle in the entropy-temperature plane~\cite{Chua},
we used an exactly solvable toy system~\cite{bakk1} to model the 
temperature dependence of the entropy of the backbone of the protein chain 
as it folds or unfolds 
with or without  guidance.
Clearly this is very schematic picture, 
but we believe it conveys the essential physics.

We assume that there is only one folding pathway leading to 
a single low-free energy folded state, to simplify the discussion. In this 
respect, then, the amino-acid chains under consideration are similar to modern 
day proteins.

The existence of a unique pathway 
means that an ordered sequence of binding events occur 
between different parts of the protein~\cite{hansen}; and if this 
particular sequence is not followed, the 
protein can not fold. 

Slightly modifying  the hierarchical 
zipper-like model proposed by Bakk 
et al.~\cite{bakk1}, let us 
assign a variable $\sigma_i=1,\ldots q$ to the 
different choices 
that can be made  at each  node, with only say 
$\sigma_i^*$ leading to a 
correct folding 
move. Then, 
the state variable at the $i$th node on 
the folding pathway of $N$ nodes~\cite{bakk1} may be written in terms of a 
Kroenecker delta as,
\be
\psi_i= \delta_{{\sigma_i},{\sigma_i^*}},\;\;\;\;\;i=1,\ldots,N \;\;\;.
\ee
\noi
The Hamiltonian is 
\be
{\cal H} = -\lambda \epsilon \sum_{i=1}^N \Psi_i - (1-\lambda) 
\epsilon \Psi_N \;\;,\label{Ham}
\ee
\noi
where
\be
\Psi_i = \prod_1^i \psi_k\;\;.\ee
\noi

For  $\lambda\neq 0$ we see that intermediate partially folded 
states also lower the energy, as would be the case under the 
guidance of chaperons~\cite{fersht}, 
by an amount $\lambda \e$ while the energy gap of the native state is given by 
$\e$.
For $\lambda=0$, this Hamiltonian
allows the protein to be in two distinct states only, native and unfolded. 
No unfolding can occur inside 
an already folded part of the protein. 
Notice that (\ref{Ham}) differs from the Bakk et al. model in the last term, 
which in our case is not multiplied by $N$; this allows the folding transition 
to shift to higher temperatures under guidence, as is experimentally 
observed~\cite{herendeen}.

The partition function for this model can be evaluated exactly. 
For a protein having a 
folding pathway of $N$ nodes, 
with  $\beta  = 1/k_BT$, $k_B$ the Boltzmann constant, and $T$ the 
temperature, 
\be 
Z =(q-1)q^{N - 1}\, 
\left( {\frac{{1 - e^{\nu N} }}{{1 - e^\nu  }}} \right) 
\,+\, e^{\beta[\lambda(N-1)+1] \epsilon }\;\;\;. 
\label{partition}
\ee
where we have defined $ \displaystyle \nu  \equiv \beta \lambda \e 
-\ln q $.  (Note that L'Hopital's rule must be used in the first term in 
case the denominator vanishes).
From the Helmholtz free energy $F =  - kT\ln Z$
one may compute  
the entropy $S=-\partial F/\partial T)_V$ and heat capacity 
$C=T\partial S/\partial T)_V$.
(In this model the free energy is independent of the volume, therefore the 
partial
derivatives above can be treated as ordinary derivatives.)
The entropy is   plotted in 
 Fig.2
 for different values of $\lambda$, $\lambda=0$ corresponding to the unguided  and $\lambda\neq 0$ to the guided cases.
The system exhibits a sharp transition (becoming first-order 
in the thermodynamic limit, i.e. 
$\lim N \rightarrow \infty$) for the unguided case, as expected from this two-
state model.  The effect of guiding is to lower the entropy, smooth the  
transition and to shift it to higher temperatures, and is similar to the effect 
of turning on a field, in the case of magnetic phase transitions.

The refrigeration cycle for the refrigerant is shown in Fig.2 in the
entropy-temperature plane. The sharp unfolding transition ($1\to
1^\prime$) analogous to the flash-evaporation in a conventional adsorption
refrigerator, takes place in the cooling chamber, at $T_c$.  The high pH
conditions we take to correspond to the $\lambda=0$ curve.  The denatured
proteins are convected towards the adsorbing walls (the walls could be coated by some lipids
to make them more hydrophobic~\cite{Martin+Russel}), warming up in the
process($1^\prime \to 2$), and adsorb ($2\to 3$). Upon adsorption
$\lambda$ is set to unity, and the entropy of the chain drops. This
happens at the (fixed) intermediate temperature $T_m$ of the adsorbing
wall, and the heat of adsorption is carried away by the water bathing the
outer wall of the compartment.  When the heat is turned on, as it
periodically is, the wall will heat up to $T_h$, along the curve $3\to 4$.  
In the process, $Q_{\rm des}$ is absorbed by the proteins.  At point 4 in
the cycle, they become free of the wall; and are now convected along $4\to
4^\prime\to 1$, back to the starting point.  This happens along a curve
with a nonzero value of $\lambda$, here chosen  to be
$\lambda=0.13$, because, as explained in item {\it (ii)} of the previous
section, we assume these proteins  remain near the hydrophobic wall as they give
off excess heat to the rock boundary washed by cooling water on the
outside, and are convected back towards the cool chamber.

The various heats of evaporation, adsorption, desorption and condensation
may, in principle, be computed from the entropy as a function of the
temperature for the various values of $\lambda$.  It is clear from the
geometry of the curves in Fig. 2 that the area to the left of the curve 
between any two points $(ij)$ on the curve may be 
obtained by doing the integral over $T$ instead, viz.,
\be
Q_{ij} = T_j S_\lambda^{(j)} - T_i S_\lambda^{(i)} 
- \int_{T_i}^{T_j} S_{\lambda}(T) dT  
\;\;\;.
\ee

The ``coefficient of performance'' of the refrigerator is given 
by~\cite{Cengel} 
\be
{\rm COP} = {Q_{\rm ev} \over Q_{\rm des} + W_g}\;\;,
\ee
where $W_g$ is a small amount of work normally needed to operate a pump.  In 
our case, this is  provided by the gravitational force driving the 
convection current.  To maximize $Q_{\rm ev}$, one must have 
essentially two-state unfolding at $T_c$, with a maximal entropy 
gap.  
On the other hand, one must clearly have 
the folding-unfolding rates, 
$Q_{\rm con}-Q_{\rm ev} < Q_{\rm des}-Q_{\rm ads}$ to respect the first 
law of thermodynamics.  For this, one needs the adsorption to take place 
at $T_m$ values that are as low as possible, while $Q_{\rm con}$ should 
also be minimized. We have chosen appropriate values of $\lambda$ in Fig. 
2 to make this plausable.

\begin{figure}
\bc
\leavevmode
\includegraphics[width=9cm,angle=0]{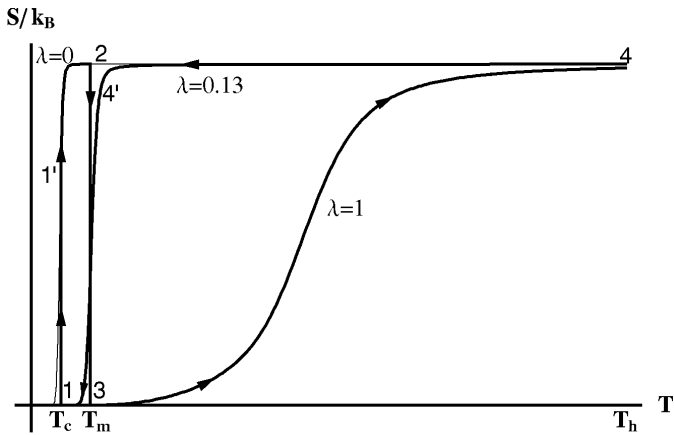}
\caption{\small{The refrigeration cycle in the entropy-temperature plane.  
The curves have been drawn for $\lambda=0,\,0.13$ and 1. For this plot  
$q=8$ and $N=10$. 
The temperature axis is in units of $\e /k_B$.}} \label{STdiagram}
\ec
\end{figure}

In order to have a crude estimate of the actual
pumping rates possible, one should note that
the entropy of denaturation of a typical protein
is about $\Delta S \sim $17J/mol-K per residue.~\cite{Privalov2,Daune} The
folding-unfolding rates range between $k \sim 10^2 $/s to $10^5 $/s~\cite{mayor,oliveberg}
whereas typical protein concentrations, namely $\rho$, are of the order $10 \mu$M~\cite{mayor}.
If one takes a cooling chamber of linear dimensions $r_c\sim 1$ cm,
one finds that the heat pumped away,  $J_{\rm ev} =  d Q_{\rm ev}/dt$ is
\be J_{\rm ev} = \rho k r_c^3 L T_c \Delta S \sim 5 \times 10^{-1}{\rm J/s}\;\;,
\ee
\noindent
for typical proteins, where $L$, the length of a protein chain, is taken 
to be $100$, $T_c=300$ K and $k=10^{-2}$/s.

On the other hand typical heat conductivities for rock are around $K_T=2-6 $
J/m-K-s~\cite{CRChandbook,AIPhandbook}. With
a temperature difference between the ambient water and the inside of the
cooling chamber being, say $T_m-T_c \sim 30$ K, and a surface area $\sim r_c^2$,
the heat conducted into
the cooling chamber from the surroundings, $J_{\rm in}=d Q_{\rm in}/dt$ will be
\be J_{\rm in}= K_T {T_m-T_c \over \ell} r_c^2\ee
where $\ell$ is the thickness of the
wall seperating the chamber from the ambient water.
With $r_c \sim$ 1 cm, and taking $K_T=4$ J/m-K-s,  one finds
$J_{\rm in}\sim (1.20/\ell) {\rm J/s}$, if $\ell$ is expressed in cm. .
The break even point where $J_{\rm in} \simeq J_{\rm ev}$ 
is around $\ell \sim 2.5$ cm.

The power needed for this system can be found from $J_{\rm des}=d Q_{\rm des}/dt$.  The
insolation at the earth surface is of the order of $W_{\rm sol}=200$ J/m$^2$-s.  Requiring
\be 
J_{s}\equiv  W_{\rm sol} r^2_{\rm b} >   J_{\rm des}=
{ J_{\rm ev} \over {\rm COP}}
\ee
where $r_{\rm b}$ is the radius of the adsorption bed receiving the solar
power and inserting numbers, one finds $r_{\rm b}$ should be of 
the order of a few tens of cm, for COP as low as 0.1.

\section{Discussion} 

We have shown above that a protein soup within a porous
rock could function as a self-regulatory refrigeration cycle and lower the
temperature of the soup within the pore, for realistic ranges of the
physical and chemical parameters.  
The efficiency of the cycle strongly
depends on the size of the entropy gaps.

It is well known that the presence of protein molecules acting as enzymes
may effect the RNA replication rates (which we have not considered) by
factors of up to $10^4$~\cite{eigen,Campbell}. Catalytic activity is a
function of the spatial structure~\cite{eigen} and therefore requires the
proteins to be in a unique folded state, whose stability also depends on
the temperature and is optimized only for a definite temperature
interval~\cite{fersht,oliveberg}. Moreover, RNA replication rates depend
non-monotonically on the temperature~\cite{Campbell} and drop off outside
a definite temperatre range.  We have shown that proteins with large
entropy gaps are able to achive temperatures in a small compartment that
are lowered relative to the ambient temperature. This vindicates our
initial assumption that this criterion could concievably have played a
role in their selection, in a environments that are too warm for the
optimal self-replication of RNA.

There is experimental evidence that the folding transition is like 
a two-state system for many 
single-domain proteins~\cite{fersht,kaya,privalov1}. 
Although the proteins
fold via a two-state folding pathway, especially at higher 
temperatures, the presence of some intermediate states might
be necessary for the folding process to find the native state. It 
has been experimentally observed~\cite{herendeen} that the 
concentration of chaperons in E. coli rises as the temperature increases, 
indicating that at high temperatures E. coli needs help in order 
to fold its proteins.

The efficiency of protein folding can be adversely affected if
partially folded proteins aggregate in order to 
reduce exposed hydrophobic residues. Molecular chaperons bind 
reversibly to these partially folded chains
preventing their aggregation and promoting their passage down the 
folding pathway. Therefore, one might  
speculate that these chaperons have taken the role of rock 
surfaces in the course of higher evolution.

It is interesting to note that there exist so called heat-shock proteins 
which are synthesized in large numbers when the temperature of the 
environment is suddenly raised above (or dropped below) those 
temperature above and below which proteins in the cells would normally 
denature, and chaperone the correct folding of other proteins in the 
cell.~\cite{arnvig} 
These proteins observed today 
under extremely hot conditions, in archeabacteria  
are  relatively small
and have very fast folding rates.\cite{Eigen1,fersht,mayor} These examples 
seem to point also in the direction of a direct correlation between 
thermal properties of the environment and protein functions, which may 
have evolved from the rudimentary function proposed here.

Other 
studies also have demonstrated that rock
surfaces may have played a selection role in prebiotic conditions, 
c.f., the selection of right handed amino acids
binding to optically active surfaces of calcite crystals with same 
chirality~\cite{hazen}.

Clearly we are making no 
claims that the present proteins are 
identical to the
end products of one particular selection mechanism. Once the temperature was
sufficiently lowered, other, more complex evolutionary pressures would come
into play. However, our calculations show that 
the proposed refrigeration cycle could very well have played an important role 
in the co-evolution of fast folding proteins with large entropy gaps 
and the RNA molecules which code them.

\end{document}